\documentclass[onecolumn]{aa} 

\usepackage{graphicx}
\usepackage{txfonts}
\usepackage[usenames, dvipsnames]{color}
\begin{document}

   \title{Differences between Doppler velocities of ions and neutral atoms in a solar prominence}


   \author{
	T. Anan
          \inst{1}
	   , K. Ichimoto
	  \inst{1,2}
          , \and
          A. Hillier
	  \inst{3}
          }

   \institute{
   		 Kwasan and Hida Observatories, Kyoto University, Gifu, Japan   \\
         \and
   		 SOLAR-C Project Office, National Astronomical Observatory of Japan, Tokyo, Japan   \\
         \and
             	 CEMPS, University of Exeter, Exeter EX4 4QF U.K.  \\
             \email{anan@kwasan.kyoto-u.ac.jp}
             }

   \date{Received XXX, 2015; accepted xxx, 2015}

 
  \abstract
   {
In astrophysical systems with partially ionized plasma the motion of ions is governed by the magnetic field while the neutral particles can only feel the magnetic field's Lorentz force indirectly through collisions with ions.
The drift in the velocity between ionized and neutral species plays a key role in modifying important physical processes like magnetic reconnection, damping of magnetohydrodynamic waves, {transport of angular momentum in plasma through the magnetic field}, and heating.
}
   %
   {
This paper investigates the differences between Doppler velocities of calcium ions and neutral hydrogen in a solar prominence to look for velocity differences between the neutral and ionized species.
   }
   {
We simultaneously observed spectra of a prominence over an active region in  H I 397 nm, H I 434 nm, Ca II 397 nm, and Ca II 854 nm using a high dispersion spectrograph of the Domeless Solar Telescope at Hida observatory, and compared the Doppler velocities, derived from the shift of the peak of the spectral lines presumably emitted from optically-thin plasma.
   }
   {
There are instances when the difference in velocities between neutral atoms and ions is significant, e.g. 1433 events ($\sim$ 3 \% of sets of compared profiles) with a difference in velocity between neutral hydrogen atoms and calcium ions greater than $3\sigma$ of the measurement error.
However, we also found significant differences between the Doppler velocities of two spectral lines emitted from the same species, and the probability density functions of velocity difference between the same species is not significantly different from those between neutral atoms and ions.
   }
   {
We interpreted the difference of Doppler velocities as a result of motions of different components in the prominence along the line of sight, rather than the decoupling of neutral atoms from plasma. 
   }

   \keywords{Sun: filaments, prominences --
                Magnetohydrodynamics (MHD) --
                Methods: observational --
                Techniques: spectroscopic
               }

   \maketitle
%

\section{Introduction}

The plasma in the solar photosphere, the solar chromosphere, the interstellar medium, and protoplanetary disks, to give just a few examples, is partially ionized.
In many of these systems, magnetic fields play an crucial role in the plasma dynamics, but they cannot directly exert a force on the neutral particles.
Therefore, the Lorentz force is indirectly exerted on the neutrals through collisional friction between the neutral and charged particles.
However, this coupling of the two fluids, and by extension the neutrals to the magnetic field, is not perfect causing the neutral particles to diffuse across the magnetic field in a process called ambipolar diffusion \citep[e.g.][]{brandenburg94}.

This diffusion of neutrals across the magnetic field is a crucial process in astrophysical systems.
It increases the rate at which magnetic fields reconnect \citep[e.g.][]{zweibel89, leake12}, damping rates of propagating magnetohydrodynamic waves \citep[e.g.][]{osterbrock61, khodachenko04}, heating rates in the solar chromosphere \citep[e.g.][]{osterbrock61, khomenko12}, and rate at which magnetic field emerges from the convection zone into the solar corona \citep[e.g.][]{arber07, leake13}.
The effect also changes the energy flux of Alfv${\rm \acute{e}}$n wave \citep{vranjes08}, the structure of slow-mode MHD shocks \citep{hillier16a}, the thermodynamic structure of quiet-Sun magnetic features \citep{cheung12}, and the structure of the solar prominences \citep{hillier10}.
Furthermore, it has significant impact on the angular momentum transport by magnetic fields in the formation and evolution of circumstellar disks and stars \citep[e.g.][]{mestel56, tomida15}.
However, direct observation of ion-neutral drift is difficult.
This is partially due to the fact that the appreciable decoupling of neutral atoms from plasma is too small to be measured on observable scales because the expected collisional coupling in many partial-ionized astrophysical plasmas is strong.

\begin{figure*}
   \centering
   \includegraphics[width=16cm]{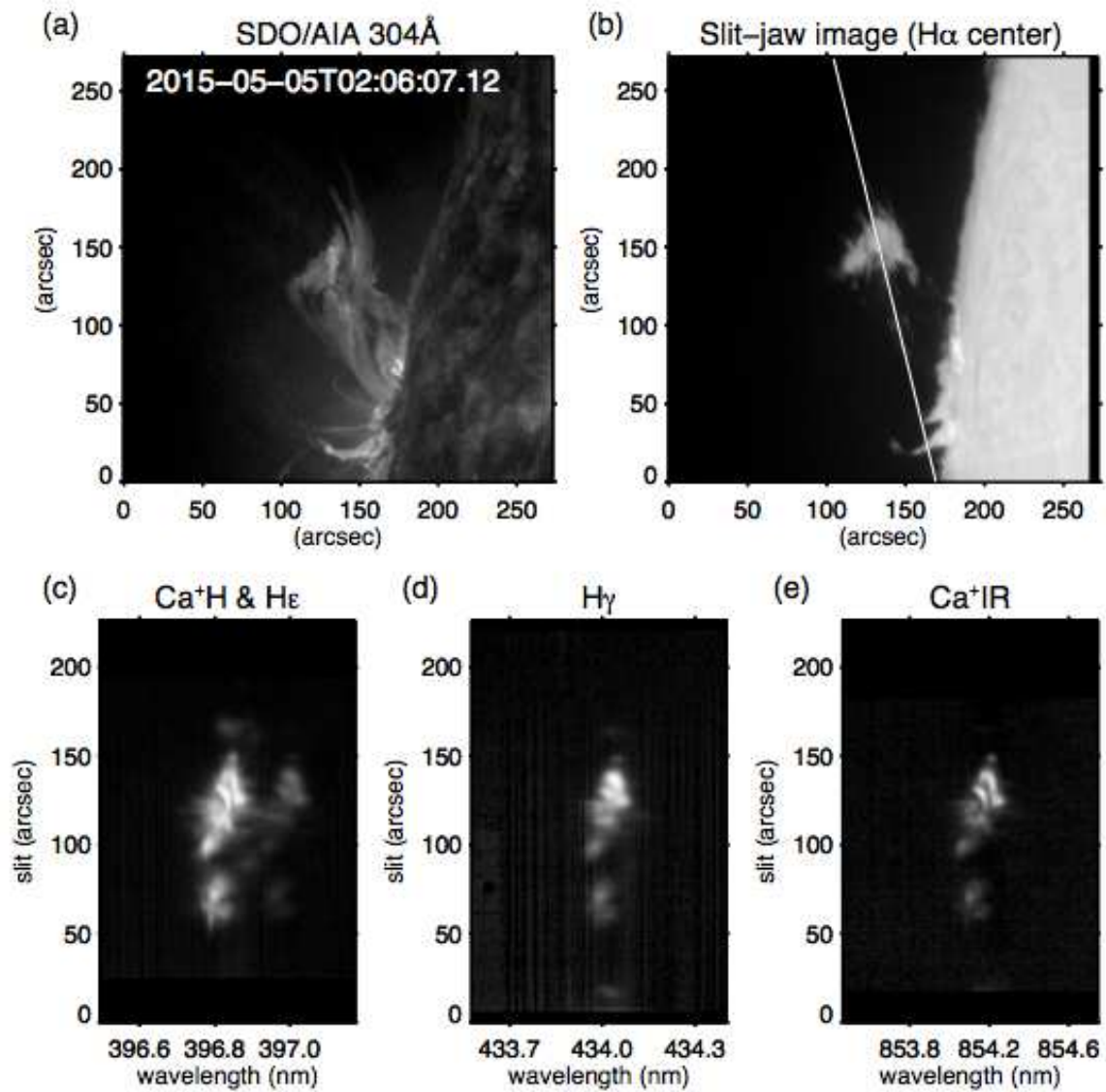}
      \caption{
		Prominence in (a) the AIA 304 \AA\, imager onboard SDO, (b) H$\alpha$ slit-jaw image, (c) spectrum including Ca$^{+}$H, and H${\rm \epsilon}$, (d) spectrum of H${\rm \gamma}$, and (e) spectrum of Ca$^{+}$IR.
		The oblique white line in the slit-jaw image shows the spectral slit.
		The temporal evolution is available online$^{1}$.
			}
         \label{Fig.obs1}
\end{figure*}

\begin{figure}
   \centering
   \includegraphics[width=8cm]{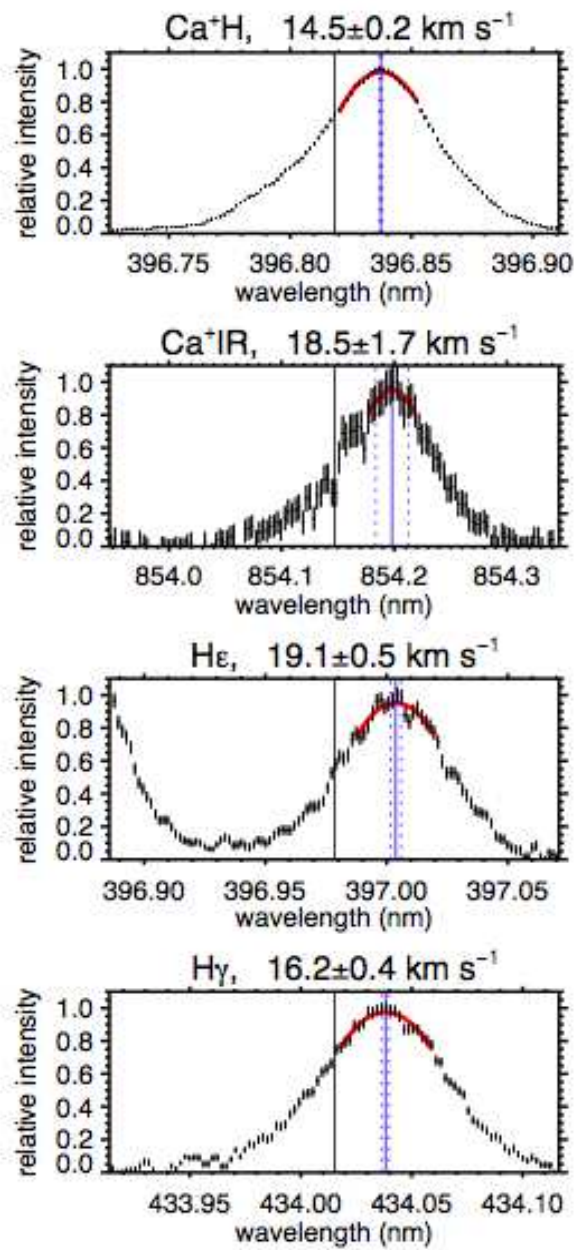}
      \caption{
      		Example of the observed profiles and fits with a parabola.
		Black symbols indicate the observed profiles in Ca$^{+}$H, Ca$^{+}$IR, H${\rm \epsilon}$, and H${\rm \gamma}$ normalized by the maximum intensity of each profile.
		The wavelength position of each peak (blue solid lines) is determined from fitting (red lines) with a parabola using interval at which the intensity is larger than 75\% of the peak intensity.
		The inferred Doppler velocities of Ca$^{+}$H, Ca$^{+}$IR, H${\rm \epsilon}$, and H${\rm \gamma}$ are $14.5 \pm 0.2 \,{\rm km\,s^{-1}}$, $18.5 \pm 1.7 \,{\rm km\,s^{-1}}$, $19.1 \pm 0.5 \,{\rm km\,s^{-1}}$, and $16.2 \pm 0.4 \,{\rm km\,s^{-1}}$, respectively.
		The measurement error is indicated in each title for the 1-sigma uncertainty, and in each plot as the blue dotted lines for the 3-sigma uncertainty.
		The minimum and maximum wavelength ranges displayed in the individual panels are $-70$ and $+70$ km s$^{-1}$, respectively.
		The black vertical solid lines denote the wavelength where the Doppler velocities are equal to zero.
      }
         \label{fig.fit}
\end{figure}

\begin{figure}
   \centering
   \includegraphics[width=5cm]{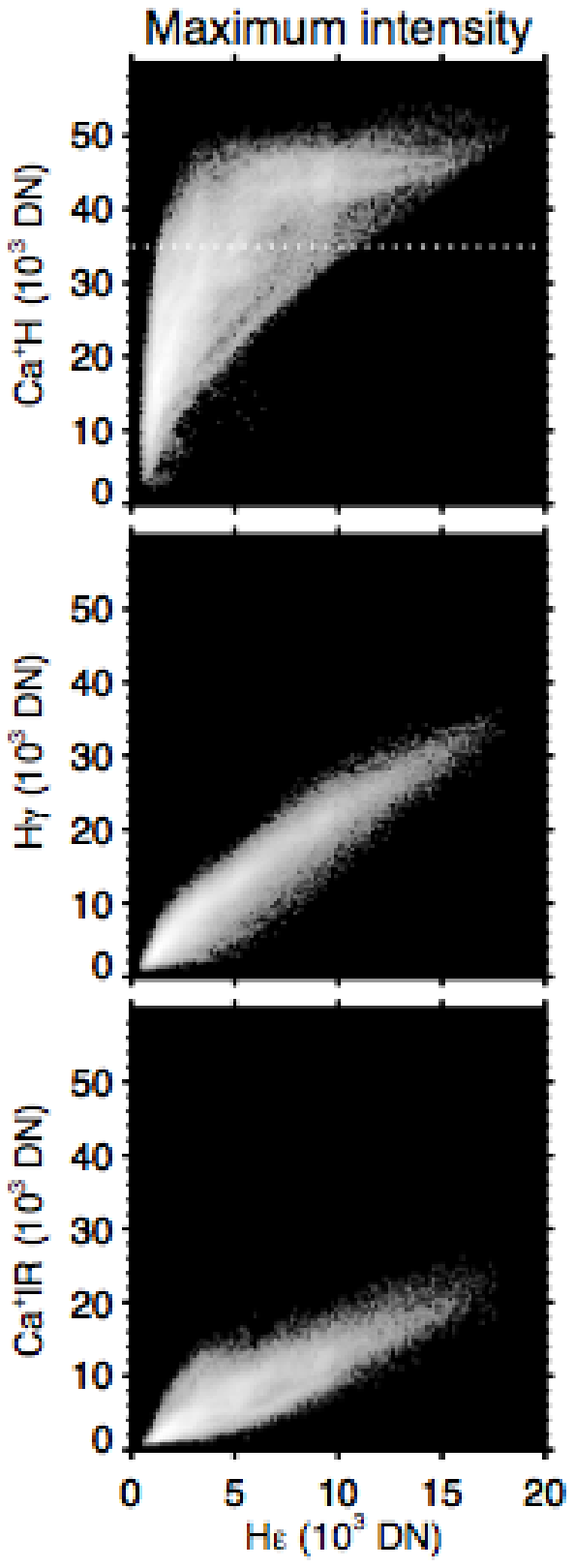}
      \caption{
		Maximum intensity of  Ca$^{+}$H vs H${\rm \epsilon}$, H${\rm \gamma}$ vs H${\rm \epsilon}$, and Ca$^{+}$IR vs H${\rm \epsilon}$.
		The white dotted line indicates a threshold in the maximum intensity of Ca$^{+}$H to determine the Ca$^{+}$H intensities that are emitted by optically-thin plasma.
      }
         \label{fig.thin}
\end{figure}

The object of our observation is a solar prominence.
Solar prominences are relatively cool ($\sim 10^4\,{\rm K}$) and dense (3 - 6 $\times10^{11}\,{\rm cm^{-3}}$) structures observed in chromospheric lines sustained in the hotter and the sparser corona above the solar limb \citep{hirayama86, tandberg-hanssen95}.
Because of their low temperature, prominences are made of partially-ionized plasma, with the ionization fraction of hydrogen characteristically 0.2 \citep{ruzdjak90, engvold90, labrosse10} in the centre of the dense prominence material.
In prominences, it has been suggested that plasma is supported against gravity by the Lorentz force \citep{kippenhahn57, kuperus74}, and neutral atoms are supported by the frictional force between them and the plasma \citep{low12}.
Prominences are dynamic structures, displaying motions of various kinds, such as turbulence \citep[e.g.][]{hillier16b}, oscillations \citep[e.g.][]{okamoto07}, and convection \citep{berger08, berger11}.

In a partially ionized system, when neglecting the electron and ion inertia terms and assuming that magnetic forces dominate, we obtain the relative velocity between ions, $v_{i}$, and neutrals, $v_{n}$ as, 
\begin{equation}
\vec{v}_{i} - \vec{v}_{n} \propto \frac{\xi}{\alpha_{n}} (\vec{J} \times \vec{B}),\nonumber
\end{equation}
where $\xi$ is the neutral fraction, $\alpha_{n}$ is the sum of collisional frequencies between the neutral and ionized species multiplied by the corresponding mass densities, $\vec{J}$ is the current and $\vec{B}$ is the magnetic field vector.

An analytical calculation \citep{gilbert02} using a simple prominence model in which the Lorentz force is balanced with the frictional force shows the relative flow of the neutral hydrogen and ionized components of $3.7 \times 10^{-3} \, {\rm km \, s^{-1}}$.
The relative flow of neutrals to ions was modeled using numerical simulations in a 2D prominence model \citep{terradas15}.
Simulations also show decoupling of neutral atoms from plasma as a result of the Rayleigh-Taylor instability in prominences \citep{khomenko14b}.

Charge exchange is another process through which ionized and neutral fluids can couple and it effectively increases the cross section of momentum transfer between neutral and ionized hydrogen by approximately a factor 2 \citep{krstic99, vranjes13}.
Charge exchange is able to increases the momentum coupling between a range of neutral atoms and ions \citep[e.g.][]{leake13b, vranjes16}, however as the process requires resonance it is naturally most common between two atoms of the same species.
\citet{terradas15} described that the relative flow of neutral hydrogen to protons in a prominence is reduced by the charge exchange interactions.




\citet{khomenko16} detected differences between ion and neutral velocities in a prominence, as shown by the difference in Doppler velocities of He I 1083 nm and Ca II 854 nm of the order of 0.1 km s$^{-1}$.
They discussed coherency of the different velocity in time and space, but they didn't compare between the same neutral species to confirm that the differences in Doppler shift observed were actually a result of ion-neutral drift.
In this paper, we present an observation of another solar prominences in two spectral lines of the neutral hydrogen and two lines of the calcium ions, in order to confirm the decoupling of neutral atoms from the plasma.
In the following sections, we describe the details of the observations (Section \ref{sec.obs}), the inference of the Doppler velocities (Section \ref{sec.mes}), the results of differences between the Doppler velocity of neutral atoms and ions (Section \ref{sec.res}), the discussions (Section \ref{sec.dis}), and finally we summarize (Section \ref{sec.sum}).  


\section{Observation and image processing}
\label{sec.obs}

A prominence over an active region NOAA12339 at the east solar limb was observed in Ca II 397 nm (Ca$^{+}$H), H I 397 nm (H${\rm \epsilon}$), H I 434 nm (H${\rm \gamma}$), and Ca II 854 nm (Ca$^{+}$IR) using the horizontal spectrograph of the Domeless Solar Telescope \citep{nakai85} at Hida observatory, Japan (Fig. \ref{Fig.obs1}).
The observation ran from 10:53 to 11:37 local standard time on 2015 May 5.
The heliocentric coordinates of the prominence at the time of observation was (N15$^{\circ}$, E90$^{\circ}$).
The prominence was dynamic, which is demonstrated by the accompanying movie\footnote{ Movie, http://www.kwasan.kyoto-u.ac.jp/$\sim$anan/shed/2016submit \_prominence\_fig1.gif \\ (a) the AIA 304 \AA\, imager onboard SDO, (b) H$\alpha$ slit-jaw image, (c) spectrum including Ca$^{+}$H, and H${\rm \epsilon}$, (d) spectrum of H${\rm \gamma}$, and (e) spectrum of Ca$^{+}$IR. The oblique white line in the slit-jaw image shows the spectral slit.}.
The mass drained to the chromosphere from the prominence along the loop-like structures with the plane-of-the-sky velocity of $\sim$ 50 km s$^{-1}$, which appear in 304 \AA\, images from Atmospheric Imaging 
Assembly (AIA) \citep{lemen12} on Solar Dynamic Observatory (SDO) \citep{pesnell12}.
In the spectra, the intensity profiles were changing quite rapidly.
%

The horizontal spectrograph can image the entire visible and near infrared solar spectrum for all the spatial points along the spectrograph slit.
Four spectral regions including 396 nm, 397 nm, 434 nm, and 854 nm are taken with three CCD cameras (Prosilica GE1650), with an exposure time of $0.6\, {\rm s}$, and with a time cadence of $1\, {\rm s}$.
One of the cameras took the spectral region which includes 396 nm and 397 nm.
The spectral samplings in 396 nm, 434 nm, and 854 nm are $17 \,{\rm m \AA \, pixel^{-1}}$, $21\,{\rm m \AA \, pixel^{-1}}$, and $32\,{\rm m \AA \, pixel^{-1}}$, respectively and the spatial sampling of spectra in 396 nm, 434 nm, and 854 nm are $0.28\,{\rm arcsec\,pixel^{-1}}$, $0.36\,{\rm arcsec\,pixel^{-1}}$ and $0.28\,{\rm arcsec\,pixel^{-1}}$, respectively.
The linear dispersion of spectra was determined by using neighboring solar lines in the background sky spectrum using the solar atlas \citep{moore66}.

When one camera starts its exposure, the camera produces a trigger signal for the other cameras to start their exposure.
The time lag of the start of an exposure between the three cameras is less than $8\,{\rm \mu s}$.
Assuming the Fried's parameter of 40 mm \citep{fried66, kawate11} and wind velocity at the turbulent latitude of 40 ${\rm m\,s^{-1}}$, the time scale of the seeing is approximately equal to $1\,{\rm ms}$, that is it is much larger than the synchronization accuracy, and thus the exposures of all three cameras can be regarded as exactly simultaneous.
Motion of the solar image on the slit caused by the local turbulence of air (seeing) was approximately 3 arcsec in amplitude during the run of the observation.
The typical spatial resolution during the observation is also about 4 arcsec, which corresponds to the Fried's parameter of about 40 mm at  600 nm.

During the observations, the zenith angle of the sun was approximately equal to $25^{\circ}$.
The solar rays pass through the atmosphere of the earth and are refracted before reaching the entrance window of the telescope.
Since the refraction angle can change with the wavelength, there may be a slight shift of images in three wavelengths.
We place the spectrograph slit on the prominence with an orientation parallel to the line connecting the prominence and the zenith, to sample exactly the same region of prominence on the slit in 396 nm, 434 nm, and 854 nm.

After dark-frame and flat-field corrections, the observed spectra are de-convolved using the point spread function of the horizontal spectrograph. 
Then we subtracted the sky spectrum, which was made from the average of 80 spectral profiles near by the prominence.
In order to compare the three spectral lines, we reduce the spatial sampling of 396 nm and 854 nm to that of 434 nm, and align the two hair lines of the spectra in 396 nm and 854 nm to those in 434 nm.
We also checked the spatial alignment of the three spectral lines using the cross-correlation between profiles of Doppler velocities of the three lines along the slit and confirmed that the differential refraction between the three lines is negligible during the observations.
Even though there was motion of the solar image on the slit caused by the seeing, the high synchronization accuracy and the negligible displacement due to the atmospheric refraction allow us to sample the same ensemble of plasma in the four spectral lines.

The point spread functions of the telescope also depend on the wavelength, $\lambda$, i.e., the theoretical size of the point spread function of the telescope is proportional to the wavelength.
However, under seeing, it is proportional to $\lambda ^{-1/5}$ \citep{roddier81}, if the exposure time is long enough compared to the seeing time scale as in our observation.
Thus the difference of the point spread function among the wavelengths is much smaller than the ideal case, and we suspect that the difference of the point spread function among the wavelengths is not a serious problem in our analysis, though we cannot absolutely exclude its small effect on our results.

The photometric accuracy of spectral measurement was evaluated from the random variation of intensity in continuum of sky component of each spectrum and it is multiplied by square root of the intensity to obtain the random error in spectral lines by assuming that the photon noise is the dominant source.
The root-mean-square photometric accuracy depends on the intensity in Ca$^{+}$H, H${\rm \epsilon}$, H${\rm \gamma}$, and Ca$^{+}$IR and varies in the range of 1 - 10\%, 1 - 20\%, 1 - 20\%, and 1 - 25\% across the respective emission lines.

\section{Doppler velocity of optically-thin plasma}
\label{sec.mes}

From the 44 min of observation, we obtained 2626 sets of spectral images of the prominence and 347677 independent sets of line profiles of the prominence.
We fit the {peak of} spectral profiles with a parabola using a wavelength interval at which the intensity is larger than 75\% of the peak intensity to determine the wavelength position of the line peak by applying the IDL routine {\it poly\_fit.pro} (Fig. \ref{fig.fit}).
We confirmed that results with different thresholds, i.e. 65 \%, 70 \%, 80 \%, and 85 \% of the peak intensity, are the qualitatively the same as those with the threshold of 75 \%.
Then, we derived the Doppler velocity of each spectral line from the shift of its peak.
The measurement error of the Doppler velocity is determined from the uncertainty of the center position of the parabola, i.e, the keyword {\it SIGMA} of the {\it poly\_fit.pro}, by which the residual error of the fitting is calculated from the random error in the spectral data.
\begin{figure*}
   \centering
   \includegraphics[width=16cm]{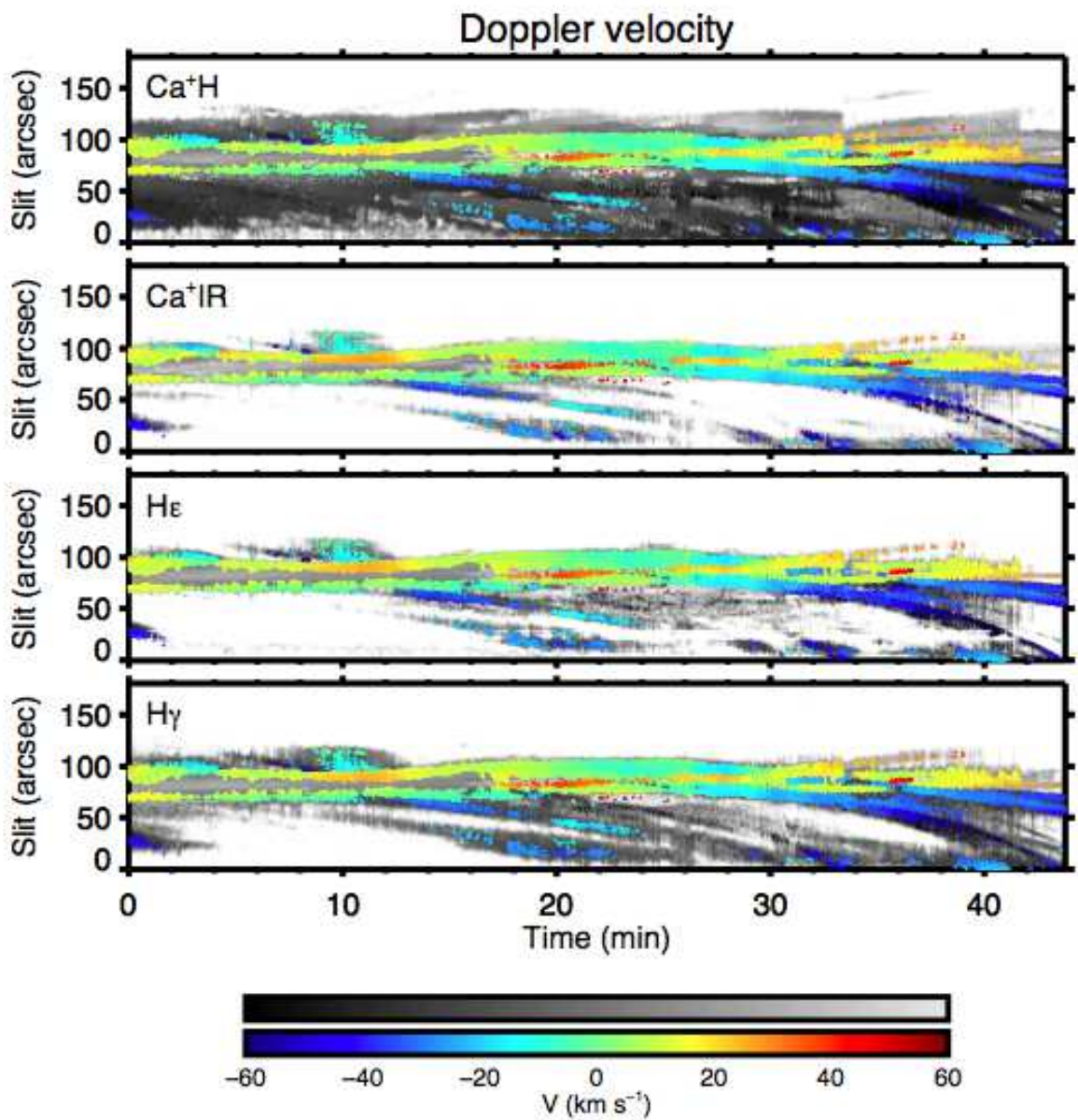}
      \caption{
      Doppler velocity of the prominence in Ca$^{+}$H, Ca$^{+}$IR, H${\rm \epsilon}$, and H${\rm \gamma}$ as functions of position along the slit and time.
      Pixels which satisfy the criteria 1), 2), 3), and 4) are shown by color scale, while the others by gray scale.
      The solar surface is located at the bottom. 
                  }
         \label{Fig.res_v}
\end{figure*}
\begin{figure*}
   \centering
   \includegraphics[width=16cm]{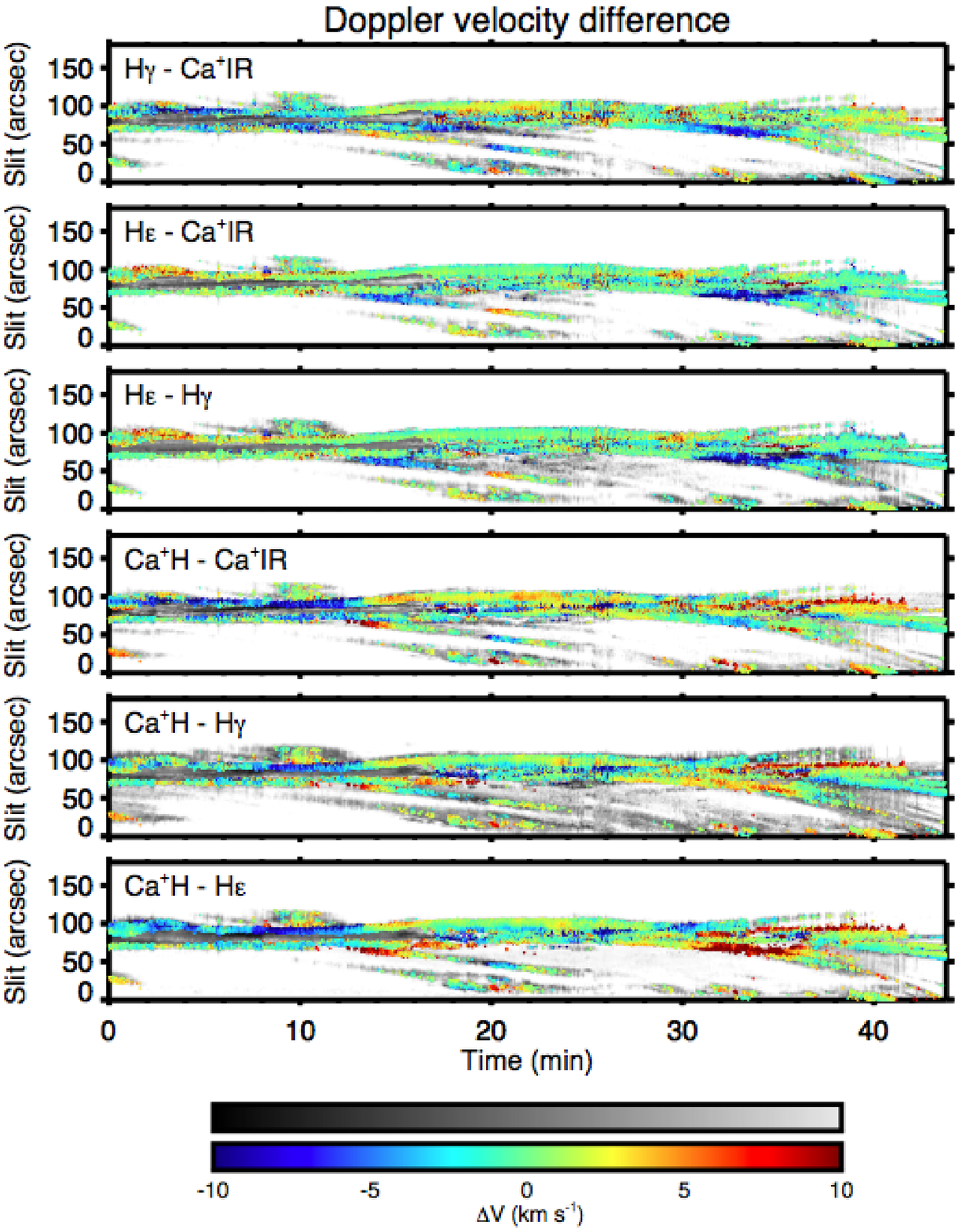}
      \caption{
      Doppler velocity differences in the prominence between pairs of spectral lines Ca$^{+}$H, Ca$^{+}$IR, H${\rm \epsilon}$, and H${\rm \gamma}$ as functions of position along the slit and time.
      Pixels which satisfy criteria 1), 2), 3), and 4) are shown by color scale, while the others by gray scale.
      The solar surface is located at the bottom. 
                  }
         \label{Fig.res_dv}
\end{figure*}
%

The observed line profiles used for the present analysis have to satisfy the following conditions:
(1) The maximum intensity of the spectral line is more than five times larger than the root-mean-square of the random noise in the background scattered spectrum; 
(2) The fitted parabola is concave downward; 
(3) The center position of the parabola is within the wavelength interval for fitting, i.e. the wavelength range where the intensity is larger than 75\% of the peak intensity, with a $99.7\%$ confidence level (3$\sigma$); and 
(4) The spectral line is emitted by {\it optically-thin} plasma as is judged by the following method.
Condition (4) is to ensure that we are able to deduce information integrated along the full depth of the emitting plasma.
In the optically thick case the radiation emitted by the plasma carries information which is dependent on the depth of the formation of the specific spectral line used for observations.
Because different optically thick spectral lines can form at very different depths, we cannot interpret difference of Doppler shifts between lines of neutral and ionized atoms as evidence of relative velocity of the different species.
Such analysis can be performed only in the optically thin case when the intensity of emission is proportional to the optical depth.
Figure \ref{fig.thin} shows a scatter plot of the maximum intensity of H${\rm \epsilon}$ vs. that of the other lines.
The Ca$^{+}$H intensity saturates due to the increase of the optical thickness while the other lines remain optically thin.
Thus, we adopt a threshold in the maximum intensity of Ca$^{+}$H as $35000$ digital number (DN) to select a data point satisfying condition (4), and to safely remove all data points in which the spectral line of Ca$^{+}$H was possibly emitted by optically thick plasma.
However, we don't adopt a threshold for the other lines, because they remain optically thin.
We note that the observational data used in this paper were not absolutely calibrated but instead the DN units are used as a measure of the intensity throughout the present paper.

Figure \ref{Fig.res_v} shows time-space diagram of the inferred Doppler velocities of Ca$^{+}$H, Ca$^{+}$IR, H${\rm \epsilon}$, and H${\rm \gamma}$ in the prominence.
The Doppler velocity of the prominence varies between $-60$ km ${\rm s^{-1}}$ and $40$ km ${\rm s^{-1}}$.
We set the zero LOS velocity as the mean position of each line over the final sample.
There are approximately 74759 pixels of the four profiles ($\sim$ 20 \% of the original data set), which satisfy the condition (1), (2), and (3) in all lines, and 43 \% of them are excluded by the (4) criterion.
Finally, the number of the pixels of the line profiles that are subject to our further analysis is $42585$.
The pixels that satisfy all conditions are shown in Fig. \ref{Fig.res_v} with the color scale.


\section{Results}
\label{sec.res}

\begin{figure}
   \centering
   \includegraphics[width=7.5cm]{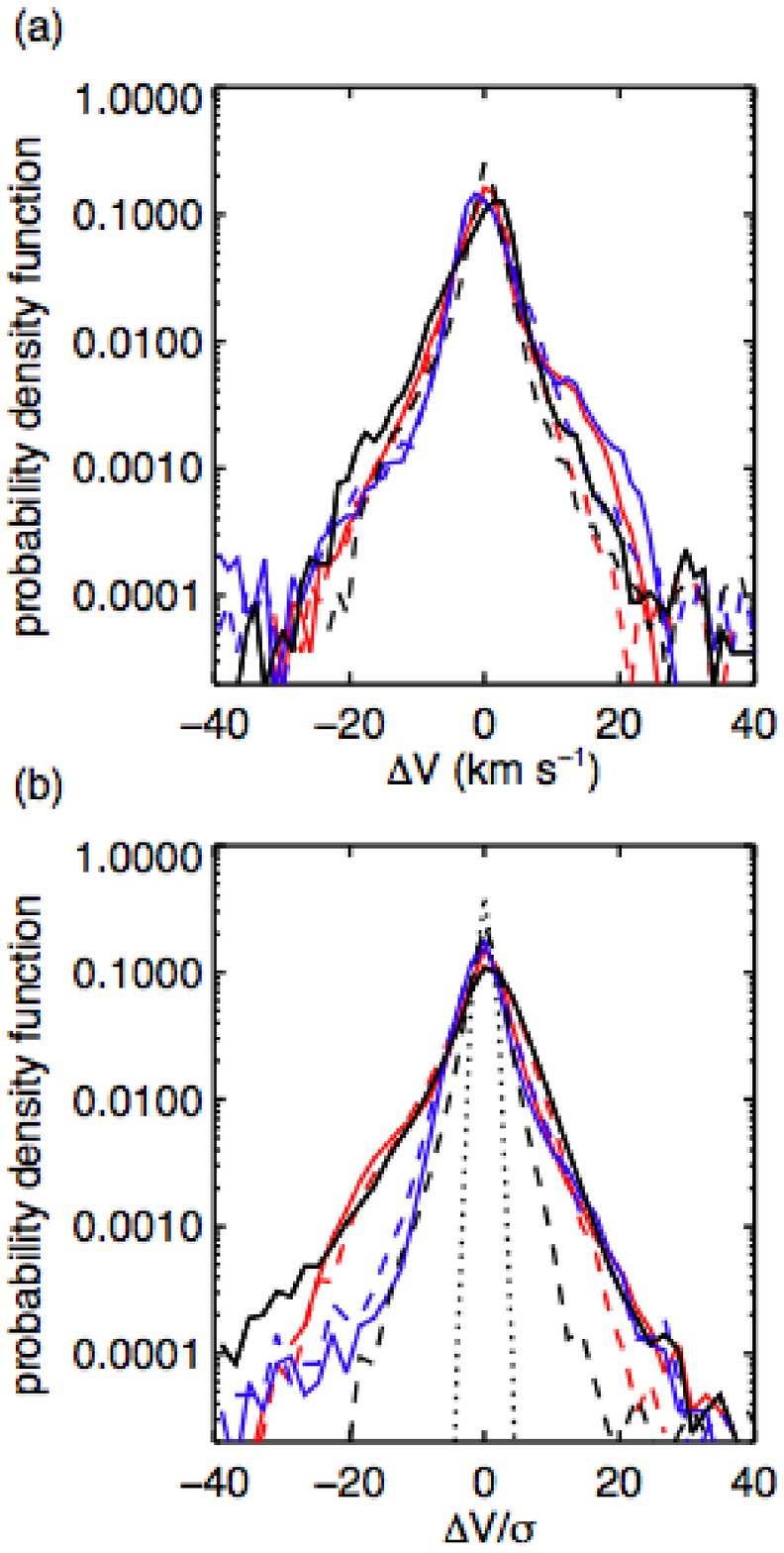}
      \caption{
      		Probability density functions of differences between the Doppler velocity of Ca$^{+}$H and H${\rm \epsilon}$ (red solid line), Ca$^{+}$H and H${\rm \gamma}$ (red dashed line), Ca$^{+}$IR and H${\rm \epsilon}$ (blue solid line), Ca$^{+}$IR and H${\rm \gamma}$ (blue dashed line), Ca$^{+}$H and Ca$^{+}$IR (black solid line), and H${\rm \epsilon}$ and H${\rm \gamma}$ (black dashed line). 
		For the functions (b), the velocity differences were normalized to the measurement errors, $\sigma$.
		The black dotted line indicates the normal distribution with a standard deviation of 1.
                  }
         \label{Fig.hist_dv_1}
\end{figure}
\begin{figure}
   \centering
   \includegraphics[width=7.5cm]{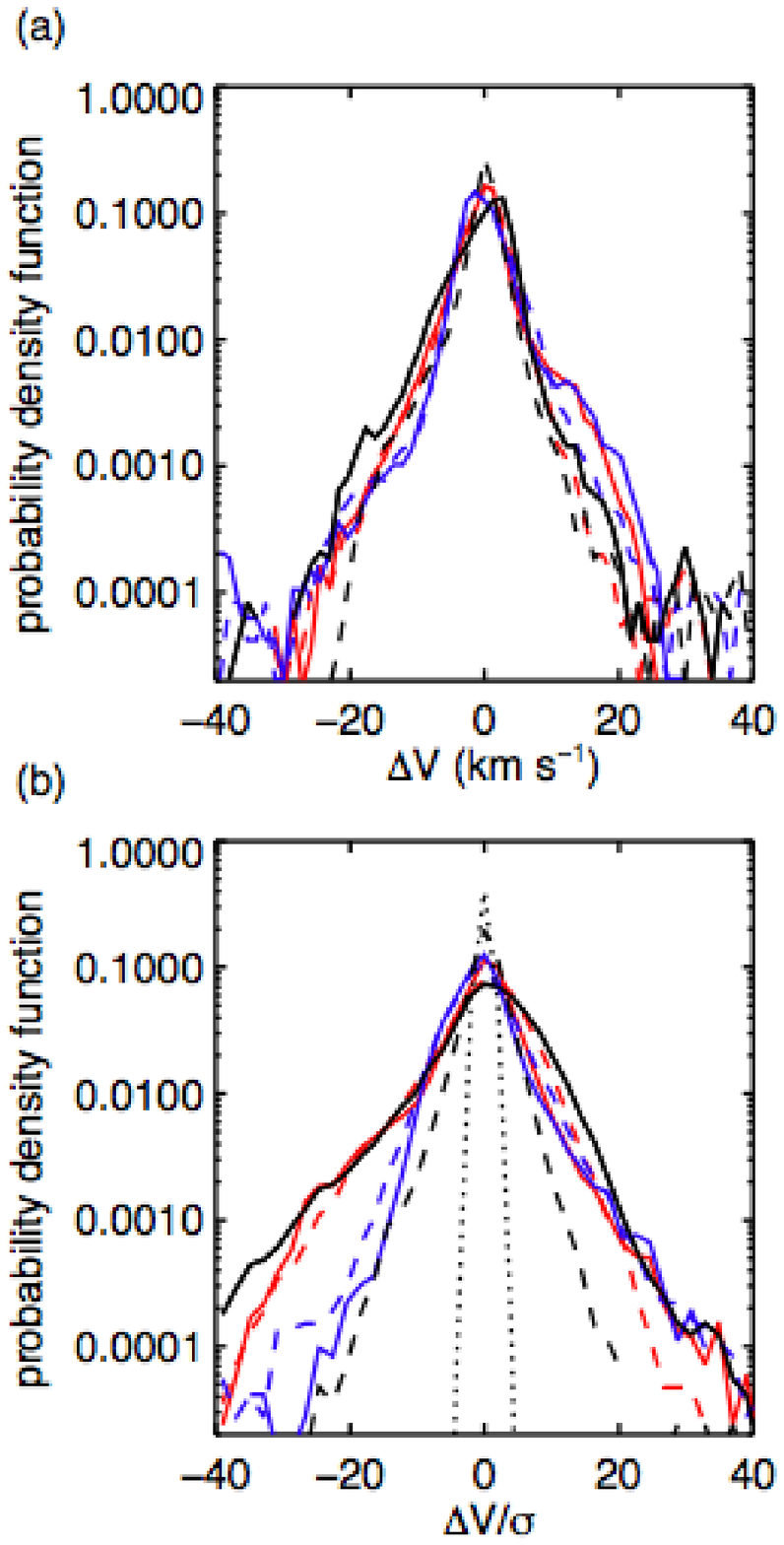}
      \caption{
      		The same as Fig. \ref{Fig.hist_dv_1}, but obtained from spectral profiles, in which the measurement error is reduced by averaging over three adjacent profiles along the slit.
                  }
         \label{Fig.hist_dv_2}
\end{figure}
%

Differences between the Doppler velocities of two spectral lines are shown as a function of space and time in Fig. \ref{Fig.res_dv}.
By comparing the Doppler velocities that are derived from the shift of the peak of the spectral lines emitted by optically-thin plasma, we find that there are instances when the difference in velocities of different lines is significant. 
For example 1433 cases ($\sim$ 3 \% of the sets of compared profiles) show a difference of velocities between neutral hydrogen atoms (H${\rm \gamma}$ and H${\rm \epsilon}$) and calcium ions (Ca$^{+}$H and Ca$^{+}$IR) greater than $3\sigma$.
Pixels with large velocity difference are located coherently in time and in space.
However, velocity differences are also significant (> 3 $\sigma$) for another 37561 pairs of spectral lines of the same species, i.e., between Ca$^{+}$H and Ca$^{+}$IR and between H${\rm \gamma}$ and H${\rm \epsilon}$.

Figure \ref{Fig.hist_dv_1} shows the probability density functions (PDFs) of the velocity difference, $\Delta V$, between two of the observed spectral lines (a) and those of velocity difference normalized by measurement error (b).
Here, the PDFs are normalized so that its integral along $\Delta V$ (a) and $\Delta V/{\rm error}$ (b) equals one.
If there is a decoupling of neutral atoms from plasma, we can expect to obtain PDFs of velocity difference between neutral atoms and ions to be wider than those of velocity differences between lines of the same species.
However, some PDFs of the velocity difference between neutral atoms and ions are narrower than either PDF of two spectral lines emitted from the same species.
The probability density function of the differences in the Doppler velocity, normalized to the measurement errors (Fig. \ref{Fig.hist_dv_1} b), display departures from the standard normal distribution (black dotted line).
If the differences were only determined by measurement errors, the distribution of the probability density function would be the same as a standard normal distribution.
This means that we have found significant Doppler velocity differences among Ca$^{+}$H, H${\rm \epsilon}$, H${\rm \gamma}$, and Ca$^{+}$IR in the prominence.

\section{Discussion}
\label{sec.dis}
In order to investigate the decoupling of neutral atoms from plasma in a solar prominence, we measured the Doppler velocities of Ca$^{+}$H, H${\rm \epsilon}$, H${\rm \gamma}$, and Ca$^{+}$IR observed simultaneously.
By comparing the Doppler velocity of neutral hydrogen and calcium ions (Fig. \ref{Fig.res_dv} and \ref{Fig.hist_dv_1}), derived from the shift of the peak of the spectral lines emitted by optically-thin plasma, we find that there is significant difference between the Doppler velocity of neutral hydrogen and calcium ions.
However, significant velocity differences are also found even between different spectral lines of the same species, for example between Ca$^{+}$H and Ca$^{+}$IR.
In this section, we discuss possible errors in the measurement and whether they cause the observed velocity differences.
Finally, we propose an interpretation of the velocity differences between spectral lines of the same species.

\subsection{Parabola fitting}
The measurement error of the Doppler velocity is determined from the uncertainty of the center position of the parabola, which correspond to the residual error of the fitting equal to the random error in spectral data.
Figure \ref{Fig.hist_dv_2} is the same as Fig. \ref{Fig.hist_dv_1}, but obtained from spectral profiles, in which the random error is reduced by averaging three adjacent profiles in the slit direction.
If the differences were only determined by the parabola fitting errors, the width of the PDFs would be $1/\sqrt{3} \sim 0.6$ times smaller than those of Fig. \ref{Fig.hist_dv_1} a.
However, the standard deviation is almost the same as that before the reduction of the random errors {(Fig. \ref{Fig.hist_dv_1} a and Fig. \ref{Fig.hist_dv_2} a)}, and the departures of the probability density function {of the velocity differences normalized to the measurement errors} from the standard normal distribution (black dotted line) are even larger than those from the single pixel profiles (compare Fig.  \ref{Fig.hist_dv_1} b and Fig. \ref{Fig.hist_dv_2} b).
Therefore, we can conclude that the velocity differences do not originate from the uncertainty of the center position of the parabola in the line fitting.

%
%

\subsection{Optical depth}
\begin{figure}
   \centering
   \includegraphics[width=8cm]{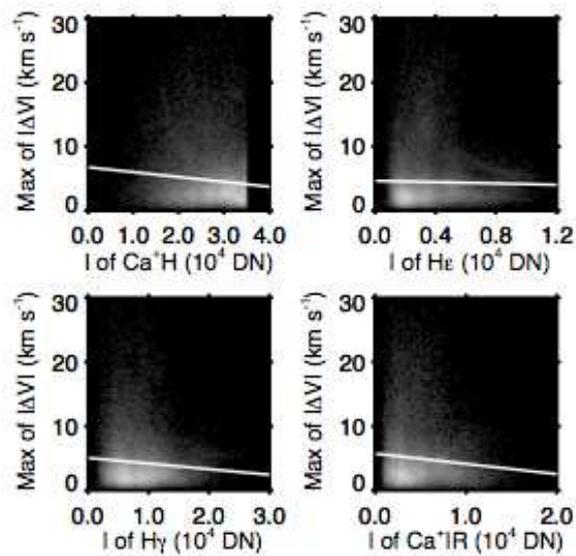}
      \caption{
      		Scatter plots of the maximum of absolute values of the Doppler velocity differences of a spectral line against the other spectral lines vs. the maximum intensity of the spectral line, and fits with a linear function (white solid lines).
                  }
         \label{Fig.hist_dv_3}
\end{figure}
If some data sets that include spectral lines emitted from optically-thick plasma remain in our selection, we are comparing the Doppler velocities of the plasma at different depth dependent on spectral lines, and the velocity difference among the four lines may increase.
Figure \ref{Fig.hist_dv_3} shows the maximum of absolute values of the Doppler velocity differences of a spectral line against the others as a function of the maximum intensity of a line.
Because the intensity of emission is proportional to the optical depth in the optically thin case, the higher intensity the higher possibility that we sample data sets including spectral lines emitted by optically-thick plasma.
But, as is shown in Fig.8, the absolute value of the Doppler velocity difference among spectral lines does not tend to be larger.
Therefore, we conclude that the optical-thickness of the plasma is not the cause of the velocity difference.

\subsection{Asymmetry of the line profile}
\begin{figure}
   \centering
   \includegraphics[width=8cm]{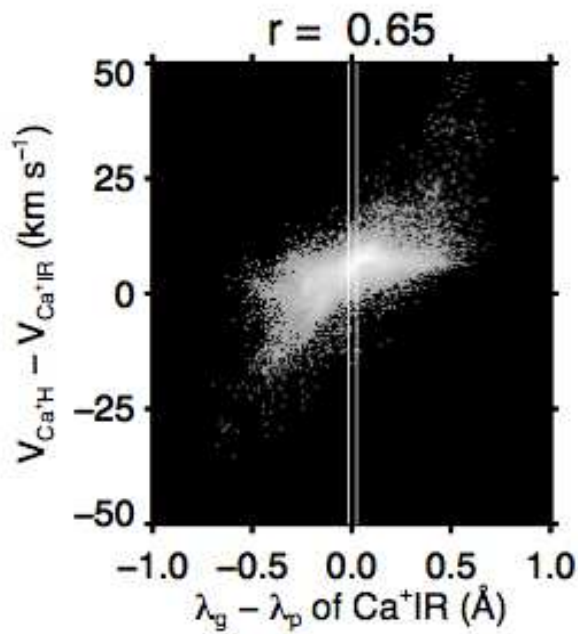}
      \caption{
		Scatter plot of the differences between the wavelength of the gravity center ($\lambda_{{\rm g}}$) and the peak ($\lambda_{{\rm p}}$) of Ca$^{+}$IR vs. the differences of Doppler velocity between Ca$^{+}$H and Ca$^{+}$IR in the pixels that satisfy all the conditions described in Section \ref{sec.mes}.
		The linear Pearson correlation coefficient, $r$, is 0.65.
		The white vertical lines indicate the threshold of $\lambda_{{\rm g}} - \lambda_{{\rm p}}$ of Ca$^{+}$IR to exclude the highly asymmetric profiles to give the absolute value of the linear Pearson correlation coefficient of all combinations of the four lines to be less than 0.3.
                  }
         \label{Fig.correlate}
\end{figure}
{For some pairs of the analyzed spectral lines, we found that differences between the wavelength of the gravity center, $\lambda_{{\rm g}}$, and the peak, $\lambda_{{\rm p}}$, of a spectral-line profile, which characterize asymmetry of the line profile, correlate with Doppler velocity differences between the spectral lines.
Figure \ref{Fig.correlate} shows an example of this strong correlation, where $\lambda_{{\rm g}} - \lambda_{{\rm p}}$ of Ca$^{+}$IR strongly correlates (linear Pearson correlation coefficient is 0.65) with Doppler velocity differences between Ca$^{+}$H and Ca$^{+}$IR, $V_{Ca^{+}H} - V_{Ca^{+}IR}$, in the pixels that satisfy all the conditions described in Section \ref{sec.mes}.}
The asymmetric profiles can be formed if there are multiple components, which have different velocities, along the line of sight or in a resolution element.
The peak wavelength may be different for different spectral lines, because sensitivities to physical parameters (e.g. density, temperature) is different for different spectral lines.
Incomplete sky subtractions can also result in asymmetric profiles.
Therefore, there is the possibility that asymmetry produces spurious Doppler velocity differences caused by the contamination of scattered light spectrum or by sampling of different plasmas.

We excluded the profiles where $|\lambda_{{\rm g}} - \lambda_{{\rm p}}|$ of a spectral-line profile is larger than a threshold.
We adopt the thresholds for the differences as $1.5 \times 10^{-3}$\,nm, $1.0 \times 10^{-2}$\,nm, and $2.0 \times 10^{-3}$\,nm in H${\rm \epsilon}$, H${\rm \gamma}$, and Ca$^{+}$IR, respectively, for the absolute value of the linear Pearson correlation coefficient of all combinations of the four lines to be less than 0.3.
For example, the correlation coefficient between $\lambda_{{\rm g}} - \lambda_{{\rm p}}$ of Ca$^{+}$IR and $V_{Ca^{+}H} - V_{Ca^{+}IR}$ becomes 0.14 after the exclusion (Fig. \ref{Fig.correlate}).
We also excluded the profiles that have two peaks within the wavelength interval for the parabola fitting.
The number of the sets of the line profiles that pass this further criteria are $1865$.

\begin{figure}
   \centering
   \includegraphics[width=7.5cm]{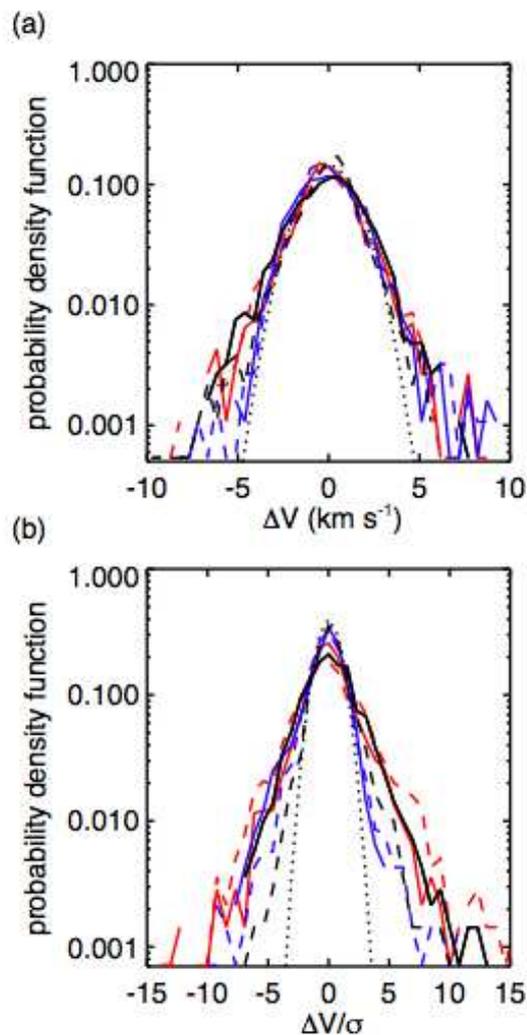}
      \caption{
      The same as Fig. \ref{Fig.hist_dv_1}, but obtained from spectral profiles, which do not have large asymmetry of the line profile or two clear peaks within the wavelength interval for the parabola fitting.
      For the functions (a), the black dotted line indicates a Gaussian distribution with a standard deviation of 1.4 km ${\rm s^{-1}}$.
      Note that the x-axes and the y-axes over smaller ranges than those of Fig. \ref{Fig.hist_dv_1}.
                 }
         \label{Fig.res_dv_gcen}
\end{figure}

Figure \ref{Fig.res_dv_gcen} is the same as Fig. \ref{Fig.hist_dv_1}, but obtained from the remaining spectral profiles.
Probability density functions of differences between the Doppler velocity of neutral hydrogen and calcium ions remarkably resemble those of the same species.
All PDFs of velocity difference are fitted with a Gaussian distribution with a standard deviation of 1.4 km ${\rm s^{-1}}$.
In section \ref{sec.res}, we described that the PDFs of velocity difference between neutral atoms and ions are not wider than those of two spectral lines emitted from the same species. 
Here, we can conclude that the asymmetry of the line profile, which affects on the measurement of the Doppler velocity in some cases, does not change the result.

We evaluated the similarity of the PDFs by using the Kolmogorov-Smirnov test, which is a method for comparing two PDFs.
From differences of their cumulative distributions, the test diagnoses the probability that a single parent distribution draws two PDFs.
We applied the test to the PDFs of velocity difference in the prominence.
The results from this test show that our PDFs are so similar that we cannot ignore the possibility of the sharing the same parent distribution for Ca$^{+}$H \& Ca$^{+}$IR, Ca$^{+}$H \& H${\rm \gamma}$, and Ca$^{+}$IR \& H${\rm \epsilon}$ (13\%).

%

%

\subsection{Interpretation}

In case that we observed the same plasma in different spectral lines we could expect that the Doppler velocity derived from the spectral lines of the same ion are the same.
However, the observations analyzed in the present paper show that significant differences exist even between spectral lines of the same species, i.e., between Ca$^{+}$H and Ca$^{+}$IR, or between H${\rm \gamma}$ and H${\rm \epsilon}$.
Moreover, the PDFs of velocity difference between neutral atoms and ions are not wider than those of two spectral lines emitted from the same species.
We concluded in the above discussions that none of the possible candidates of the error explain the distribution of the velocity differences.

Analyzing the spectral line emitted by optically-thin parts of the prominence, we measured the Doppler velocity of the emitting plasma integrated along the full depth of the prominence.
Moreover, we assumed that only the dominant component is sampled along the line-of-sight for each line, because the Doppler velocity is determined from the peak of the spectral line.
However, the prominence is not necessarily optically thin for the incident radiation that excites the atoms or the ions.
Non-LTE models of prominences show that the complex absorption and emission of the Lyman lines have to be taken into account to explain the observed emission lines from prominences \citep{gunar08,gunar10,schwartz15}, and it is shown that Ly${\rm \beta}$ is the key for the formation of H${\rm \alpha}$ for example \citep{gunar12}.
In our case, Ly${\rm \delta}$ and Ly${\rm \zeta}$ are the main spectral lines that excite neutral hydrogen to the upper state of H${\rm \gamma}$ and H${\rm \epsilon}$, respectively.
The optical depths of prominences in Ly${\rm \delta}$ and Ly${\rm \zeta}$ are generally larger than those in H${\rm \gamma}$ and H${\rm \epsilon}$, which we observed.
{The path of the incident radiation is also different from that of the observed radiation, as the incident radiation is emitted mainly from the solar surface.}
Therefore, the distribution of the neutral excited hydrogen atoms emitting H${\rm \gamma}$ depends on the radiation field of Ly${\rm \delta}$ and the distribution of atoms emitting H${\rm \epsilon}$ depends on the radiation field of Ly${\rm \zeta}$.
This means that the ratio between the number density of neutral hydrogen atoms emitting H${\rm \gamma}$ and H${\rm \epsilon}$ is not constant throughout the observed prominence (Gun${\rm \acute{a}}$r, private communication).
Such difference may result in a different rate of emission of the H${\rm \gamma}$ and H${\rm \epsilon}$ lines at different positions along a line of sight.
The same arguments as are presented here for the hydrogen lines can be expected to be valid also for calcium lines.
Hence, the dominant component in a emission line is not necessarily the same as that of another line in the same species.

High-resolution images and a radiative transfer visualization technique demonstrate that thin threads constitute the fundamental structure of the prominences \citep{lin08, gunar15a, gunar15b}.
The inferred significant velocity differences between two spectral lines of the same species may {be attributed to velocity differences} between the threads along the line-of-sight \citep[e.g.][]{gunar08}.
Because the PDFs of the velocity differences between spectral lines of calcium ions and neutral hydrogen remarkably resemble the velocity differences for the same species, 
these differences also may show the velocity differences among threads, rather than the decoupling of neutral atoms from plasma. 
Charge exchange may contribute to the reduction of the relative flow velocity of neutrals and ions in prominences, but as with collisional coupling it would not provide an explanation for the velocity difference between lines from the same species.

\section{Summary}
\label{sec.sum}

The drift in the velocity between ionized and neutral species plays a key role in modifying important physical processes like magnetic reconnection, damping of magnetohydrodynamic waves, transport of angular momenta in the formation and evolution of stars and disks, and heating in the solar chromosphere.
Khomenko et al. (2016) detected the differences in the Doppler velocity of He I 1083 nm and Ca II 854 nm in prominences, and interpreted as the drift of neutral atoms from ions.

This paper presents an analysis of the difference between the Doppler velocities of ions and neutral atoms in an active region prominence.
We use observations of the spectral lines of H${\rm \epsilon}$, H${\rm \gamma}$, Ca$^{+}$H, and Ca$^{+}$IR obtained by a high dispersion spectrograph of the Domeless Solar Telescope at Hida observatory.
We compared the Doppler velocities, derived from the shift of the peak of the spectral lines emitted by optically-thin plasma.
There are instances when the difference between the velocities of neutral atoms and ions is significant, e.g. 1433 events ($\sim$ 3 \% of sets of compared profiles) {show} a difference in the velocity between neutral hydrogen and calcium ions greater than $3\sigma$.
However, velocity differences are also significant between different spectral lines of the same species, and their PDFs remarkably resemble those of the velocity difference between neutral atoms and ions.
{We} interpreted the difference of Doppler velocities observed in different spectral lines as a result of motions of different components in the prominence in a resolution element, rather than the decoupling of neutral atoms from plasma.
{In our interpretation, different spectral lines sample different components in the prominence because the optical depth and the path of the incident radiation, which excites the atoms into the upper levels that emit the observed spectral lines, are different among the observed spectral lines.
}

Electric fields act on neutral atoms that decouple from plasma and move across the magnetic field.
If we assume a magnetic field strength of 200 G in the prominence over the active region, the electric field experienced by neutral atoms moving across the magnetic field is 0.56 V cm$^{-1}$ with a speed of 2.8 km s$^{-1}$, which corresponds to twice of the standard deviation of the Gauss distribution in Fig. \ref{Fig.res_dv_gcen} (a).
\citet{anan14} observed the full Stokes spectra of the Paschen series of neutral hydrogen in chromospheric jets, and obtained upper limits for possible electric fields of 0.3 V cm$^{-1}$ using magnetic field strength of 200 G.
Thus, there is possibility that the spectro-polarimetric observations allow us to investigate directly whether or not the decoupling of neutral atoms from the plasma cause the inferred Doppler velocity differences between calcium ions and neutral hydrogen.
In contrast, we cannot determine the decoupling of neutral atoms from the observational data of this study.

\begin{acknowledgements}
This work was supported by a Grant-in-Aid for Scientific Research (No. 22244013, P.I. K. Ichimoto; No. 15K17609, P.I. T. Anan; No. 16H01177, P.I. T. Anan) from the Ministry of Education, Culture, Sports, Science and Technology of Japan.
A.H. is supported by his STFC Emest Rutherford Fellowship grant number ST/L00397X/2.
We thank all the staff and students of Kwasan and Hida Observatory, especially Dr. H. Isobe, Mr. S. Ueno, and Dr. S. Nagata.
We also want to thank Dr. S. Gun${\rm \acute{a}}$r for his helpful comments and corrections after careful reviewing the manuscript.
We are grateful to Mr. J. Maekawa of Carl Zeiss Corp. for outstanding work for the maintenance of the Domeless Solar Telescope for 36 years.

\end{acknowledgements}

%

\begin{thebibliography}{}

  \bibitem[Anan et al. (2014)]{anan14} Anan, T., Casini, R. \& Ichimoto, K. 2014,
	ApJ, 786, 94

  \bibitem[Arber et al. (2007)]{arber07} Arber, T. D., Haynes, M. \& Leake, J. E. 2007, 
  	ApJ, 666, 541

  \bibitem[Berger et al. (2008)]{berger08} Berger, T. E., Shine, R. A., Slater, G. L., Terbell, T. D., Title, A. M., Okamoto, T. J., Ichimoto, K., Katsukawa, Y., Suematsu, Y., Tsuneta, S., Lites, B. W. \& Shimizu, T. 2008,
  	ApJ, 676, 89

  \bibitem[Berger et al. (2011)]{berger11} Berger, T. E., Testa, P., Hillier, A., Boerner, P., Low, B. C., Shibata, K., Schrijver, C., Tarbell, T. \& Title, A. 2011,
  	Nature, 472, 197
  
  \bibitem[Brandenburg \& Zweibel (1994)]{brandenburg94} Brandenburg, A. \& Zweibel, E. G. 1994,
	ApJ, 427, L91

  \bibitem[Cheung \& Cameron (2012)]{cheung12} Cheung, M. C. M. \& Cameron, R. H. 2012, 
  	ApJ, 750, 6

  \bibitem[Diaz et al. (2014)]{diaz14} D${\rm \acute{i}}$az, A. J., Khomenko, E. \& Collados, M. 2014, 
  	A\&A, 564, 97

  \bibitem[Engvold et al. (1990)]{engvold90} Engvold, O., Hirayama, T., Leroy, J. L., Priest, E. R. \& Tandberg-Hanssen, E. 1990, 
  	LNP, 363, 294

  \bibitem[Fried (1966)]{fried66} Fried, D. L. 1966, 
  	JOSA, 56, 1372  

  \bibitem[Gilbert et al. (2002)]{gilbert02} Gilbert, H. R., Hansteen, V., H., \& Holzer, T. E. 2002, 
	ApJ, 577, 464

  \bibitem[Gun\'{a}r et al. (2008)]{gunar08} Gun${\rm \acute{a}}$r, S., Heinzel, P., Anzer, U. \& Schmieder, B. 2008, 
  	A\&A, 490, 307

  \bibitem[Gun\'{a}r et al. (2010)]{gunar10} Gun${\rm \acute{a}}$r, S., Schwartz, P., Schmieder, B., Heinzel, P. \& Anzer, U. 2010, 
  	A\&A, 514, 43

  \bibitem[Gun\'{a}r et al. (2012)]{gunar12} Gun${\rm \acute{a}}$r, S., Mein, P., Schmieder, B., Heinzel, P. \& Mein, N. 2012, 
  	A\&A, 543, 93

  \bibitem[Gun\'{a}r \& Mackay (2015a)]{gunar15a} Gun${\rm \acute{a}}$r, S. \& Mackay, D. H. 2015a, 
  	ApJ, 803, 64

  \bibitem[Gun\'{a}r \& Mackay (2015b)]{gunar15b} Gun${\rm \acute{a}}$r, S. \& Mackay, D. H. 2015b, 
  	ApJ, 812, 93

  \bibitem[Hillier et al. (2010)]{hillier10} Hillier, A., Shibata, K. \& Isobe, H. 2010,
  	PASJ, 62, 1231

  \bibitem[Hillier et al. (2016)]{hillier16a} Hillier, A., Takasao, S. \& Nakamura, N. 2016,
  	A\&A, 591, 112

  \bibitem[Hillier et al. (2017)]{hillier16b} Hillier, A., Matsumoto, T. \& Ichimoto, K. 2017,
  	A\&A, 597, A111

  \bibitem[Hirayama(1986)]{hirayama86} Hirayama. T. 1986, 
	in Coronal and Prominence Plasmas, ed. A. I. Poland (Washington: NASA), 149

  \bibitem[Kawate et al. (2011)]{kawate11} Kawate, T., Hanaoka, Y., Ichimoto, K. \& Miura, N. 2011, 
  	MNRAS, 416, 2154

  \bibitem[Khodachenko et al. (2004)]{khodachenko04} Khodachenko, M. L., Arber, T. D., Rucker, H. O. \& Hanslmeier, A. 2004, 
  	A\&A, 422, 1073

  \bibitem[Khomenko \& Collados (2012)]{khomenko12} Khomenko, E. \& Collados, M. 2012, 
  	ApJ, 747, 87

  \bibitem[Khomenko et al. (2014a)]{khomenko14a} Khomenko, E., Collados, M., D${\rm \acute{i}}$az, A. \& Vitas, N. 2014,
  	Physics of Plasmas, 21, 092901

  \bibitem[Khomenko et al. (2014b)]{khomenko14b} Khomenko, E., D${\rm \acute{i}}$az, A., de Vicente, A., Collados, M. \& Luna, M. 2014,
  	A\&A, 565, 45

  \bibitem[Khomenko et al. (2016)]{khomenko16} Khomenko, E., Collados, M. \& D${\rm \acute{i}}$az, A. 2016, 
  	ApJ, 823, 132

  \bibitem[Kippenhahn \& Schl\"{u}ter (1957)]{kippenhahn57} Kippenhahn, R. \& Schl${\rm \ddot{u}}$ter, A. 1957, 
	Zeitschrift f$\ddot{u}$r Astrophysik, 43, 36

  \bibitem[Krstic \& Schultz (1999)]{krstic99} Krstic, P. S. \& Schultz, D. R. 1999,
  	J. Phys. B: At. Mol. Opt. Phys., 32, 3485

  \bibitem[Kuperus \& Raadu (1974)]{kuperus74} Kuperus, M. \& Raadu, M. A. 1974, 
	A\&A, 31, 189

  \bibitem[Labrosse et al. (2010)]{labrosse10} Labrosse, N., Heinzel, P., Vial, J.-C., Kucera, T., Parenti, S., Gun${\rm \acute{a}}$r, S., Schmieder, B. \& Kilper, G. 2010, 
	SSRv, 151, 243
	
  \bibitem[Lemen et al. (2012)]{lemen12} Lemen, J. R., et al. 2012, 
  	SoPh, 275, 17

  \bibitem[Leake et al. (2012)]{leake12} Leake, J. E., Vyacheslav, S. L., Linton, M. G. \& Meier, 2012, 
  	ApJ, 760, 109

  \bibitem[Leake \& Linton (2013a)]{leake13} Leake, J. E. \& Linton, M. G. 2013a, 
  	ApJ, 764, 54

  \bibitem[Leake \& Linton (2013b)]{leake13b} Leake, J. E. Lukin, V. S. \& Linton, M. G. 2013b, 
  	PhPl, 20, 1202

  \bibitem[Lin et al. (2008)]{lin08} Lin, Y., Martin, S. F. \& Engvold, O. 2008, 
  	ASPC, 383, 235

  \bibitem[Low et al. (2012)]{low12} Low, B. C., Berger, T., Casini, R. \& Liu, W. 2012, 
  	ApJ, 755, 34

  \bibitem[Mestel \& Spitzer (1956)]{mestel56} Mestel, L. \& Spitzer, L. Jr. 1956, 
  	MNRAS, 116, 503

   \bibitem[Moore et al. (1966)]{moore66} Moore, C. E., Minnaert, M. G. J. \& Houtgast, J. 1966, 
   	THE SOLAR SPECTRUM 2935\AA \,to 8770\AA

  \bibitem[Nakai \& Hattori (1985)]{nakai85} Nakai, Y., \& Hattori, A. 1985, 
  	Memoirs of the Faculty of Science, Kyoto University, 36, 385

  \bibitem[Okamoto et al. (2007)]{okamoto07} Okamoto, T. J., Tsuneta, S., Berger, T. E., Ichimoto, K., Katsukawa, Y., Lites, B. W., Nagata, S., Shibata, K., Shimizu, T., Shine, R. A., Suematsu, Y., Tarbell, T. D. \& Title, A. M. 2007, 
  	Science, 318, 1577

  \bibitem[Osterbrock(1961)]{osterbrock61} Osterbrock, D. E. 1961, 
  	ApJ, 134, 347

  \bibitem[Pesnell et al. (2012)]{pesnell12} Pesnell, W. D., Thompson, B. J. \& Chamberlin, P. C. 2012, 
  	SoPh, 275, 3

  \bibitem[Schwartz et al. (2015)]{schwartz15} Schwartz, P., Gun${\rm \acute{a}}$r, S. \& Curdt, W. 2015, 
  	A\&A, 577, 92
	
  \bibitem[Tandberg-Hanssen(1995)]{tandberg-hanssen95} Tandberg-Hanssen, E. 1995, 
	The Nature of Solar Prominences

  \bibitem[Terradas et al. (2015)]{terradas15} Terradas, J., Soler, R., Oliver, R. \& Ballester, J. L. 2015, 
  	ApJ, 802, 28

  \bibitem[Tomida et al. (2015)]{tomida15} Tomida, K., Okuzumi, S. \& Machida, M. N. 2015, 
  	ApJ, 801, 117
	
  \bibitem[Roddier (1981)]{roddier81} Roddier, F. 1981, 
  	PrOpt, 19, 281

  \bibitem[Ruzdjak \& Tandberg-Hanssen (1989)]{ruzdjak90} Ru{${\rm \breve{z}}$}djak, V. \& Tandberg-Hanssen, E. 1990, 
	Lecture notes in physics, 363, 317
	
  \bibitem[Vranjes et al. (2008)]{vranjes08} Vranjes, J., Poedts, S., Pandey, B. P. \& de Pontieu, B. 2008, 
	A\&A, 478, 553

  \bibitem[Vranjes \& Krstic (2013)]{vranjes13} Vranjes, J. \& Krstic, P. S. 2013, 
	A\&A, 554, A22

  \bibitem[Vranjes et al. (2016)]{vranjes16} Vranjes, J., Kono, M. \& Luna, M. 2016, 
	MNRAS, 455, 3901

  \bibitem[Zaqarashvili et al.(2011)]{zaqarashvili11} Zaqarashvili, T. V., Khodachenko, M. L. \& Rucker, H. O. 2011, 
	A\&A, 534, 93

  \bibitem[Zweibel(1989)]{zweibel89} Zweibel, E. G. 1989, 
	ApJ, 340, 550


\end{thebibliography}
%

\end{document}